\documentclass[reprint,
superscriptaddress,
showpacs,
nofootinbib,
prl,
aps,
floatfix,
]{revtex4-1}

\usepackage{amsmath,amssymb,amsfonts,amsthm}
\usepackage[english]{babel}
\usepackage{url}
\usepackage{bm}
\usepackage[latin1]{inputenc}
\usepackage{graphicx,rotating}
\usepackage[colorlinks=true,linkcolor=black, citecolor=blue]{hyperref}
\usepackage{csvsimple}
\usepackage{booktabs}
\usepackage{multirow, array}

\usepackage{natbib}

\begin{document}
\title{Radiative corrections to Gamow-Teller decays}
\author{L. Hayen}
\email[Corresponding author: ]{leendert.hayen@kuleuven.be}
\affiliation{Instituut voor Kern- en Stralingsfysica, KU Leuven, Celestijnenlaan 200D, B-3001 Leuven, Belgium}

\author{N. Severijns}
\affiliation{Instituut voor Kern- en Stralingsfysica, KU Leuven, Celestijnenlaan 200D, B-3001 Leuven, Belgium}

\date{\today}
\begin{abstract}
Radiative corrections in the electroweak sector constitute an essential component in the ability to disentangle Beyond Standard Model physics from experimental data. This is particularly relevant for strongly bound systems undergoing weak decays such as nuclear $\beta$ decay, where its contribution to top-row CKM unitarity tests is essential. In this Letter we note the need for an additional radiative correction to the Gamow-Teller form factor in allowed decays. It concerns a combination of electrostatic final state interactions and QCD-induced currents. We review the basic derivation and report analytical results. Due to differences in their theoretical treatment in the literature, effects on the neutron and mirror systems are distinct. Significant consequences appear for a comparison of the former with lattice QCD, while changes occur in the $|V_{ud}|$ extraction in the latter. We discuss new limits on right-handed currents and provide a new value for $|V_{ud}|$ from mirror decays.
\end{abstract}

\maketitle

An improved understanding of radiative corrections was at the heart of electroweak unification \cite{Berman1962, Sirlin2013} and continues to be of paramount importance as experimental results provide stringent limits on TeV-scale Beyond Standard Model Physics (BSM) \cite{Holstein2014, Cirigliano2013b}. Some of the most stringent limits on new high-scale physics come from precision studies of (nuclear) $\beta$ decay, in particular in studying the Cabibbo-Kobayashi-Maskawa (CKM) quark-mixing matrix \cite{Donoghue1992, Towner2010a}. 
Unitarity tests constitute a powerful consistency check, and tremendous effort has gone towards its evaluation since its inception \cite{Czarnecki2004}. The top-row unitarity requirement is dominated by the up-down mixing-matrix element, $V_{ud}$, for which the most precise value currently comes from superallowed nuclear $\beta$ decay \cite{Hardy2015}. Due to ever-increasing experimental precision on the lifetime and $\beta$-asymmetry over the past decades, the study of the neutron is rapidly becoming an equally important component in its determination \cite{Pocanic2017, Brown2018, Markisch2018}. Due to tremendous progress in the lattice QCD (LQCD) community in recent years \cite{Chang2018, Gupta2018}, comparison of the axial vector coupling with experiment yields competitive bounds on exotic couplings. Superallowed transitions in mirror decays have additionally demonstrated great potential and map out a complete nuclear data set \cite{Severijns2008, Naviliat-Cuncic2009}.

Extraction of precise results relies, however, on theoretical input of at least equal measure. As such, detailed studies were performed over the years by several authors \cite{Wilkinson1982, Garcia2001, Czarnecki2004, Ando2004, Garcia-Luna2006, Gudkov2006, Abele2008, Dubbers2011, Towner2010a, Sirlin2013}. In all of these, particular attention was given to final state interactions and QCD-induced influences. Interestingly, however, the classical work for the neutron by Wilkinson \cite{Wilkinson1982} and subsequent efforts appear to have neglected a part in the combination of these effects, despite sizable deviations shown several years earlier \cite{Bottino1974, Holstein1974b, Holstein1979}. For simplicity, we focus first on the neutron and later generalize to $T=1/2$ mirror decays.

The master equation for the decay of the neutron is
\begin{equation}
    |V_{ud}|^2\tau_n\left(f_V+3f_A\lambda^2\right) = \frac{2\pi^3}{G_F^2m_e^5g_V^2}\frac{1}{1+RC}
    \label{eq:general_tiple_relation}
\end{equation}
where $\tau_n$ is the neutron lifetime, $G_F \approx 10^{-5}/m_p^2$ is the Fermi coupling constant, $m_e$ is the electron mass, $\lambda \equiv g_A/g_V$ is the ratio of axial and vector coupling constants and $f_{V/A}$ their respective phase space integrals, and $RC$ represents inner and outer radiative corrections \cite{Towner2010a}. In writing it in this way, all radiative corrections to the relevant order are assumed to be absorbed into $RC$. Any deviation of $g_A$ from unity arises then from QCD.

Following Wilkinson \cite{Wilkinson1982}, the difference between the phase space integrals for vector and axial terms, $f_{V,A}$, has been assumed to be only a few parts in $10^6$ and therefore ignored. From the conclusions by Bottino \emph{et al.} and Holstein \cite{Bottino1974, Holstein1974b, Holstein1979}, however, corrections arise specifically to $f_A$ of $\mathcal{O}(\alpha Z / MR)$, where $\alpha$ is the fine-structure constant, $Z$ is the proton number of the final state, $M$ is the nuclear mass and $R$ its charge radius. The emergence of these terms is non-trivial, and great care must be taken when comparing to experimental results. We shall see that the combination of these two facts leads to changes in comparing experimental $\lambda$ measurements to LQCD and the $|V_{ud}|$ extraction in mirror systems.

We take back a few steps, and briefly review the influence of the strong interaction on the weak nuclear current. Under the assumption of a purely $V$-$A$ weak interaction, we introduce the well-known result for a $J=1/2 \to J'=1/2$ allowed transition
\begin{align}
    &\langle f(p_f) | V_\mu + A_\mu | i(p_i) \rangle = \nonumber\\
    &\bar{u}_p(p_f) \left\{g_V \gamma_\mu + i \frac{g_M}{2M}\sigma_{\mu\nu}q^\nu + \frac{g_S}{2M}q_\mu +\right. \nonumber \\
    &\left. g_A \gamma_\mu\gamma^5 + i\frac{g_T}{2M}\sigma_{\mu\nu}q^\nu\gamma^5 + \frac{g_P}{2M}q_\mu\gamma^5\right\}u_n(p_i)
    \label{eq:nuclear_current}
\end{align}
where $q=p_f-p_i$ is the momentum transfer, all $g_i$ are functions of $q^2$, and the additional terms represent the weak magnetism ($g_M$), induced scalar ($g_S$), induced tensor ($g_T$) and induced pseudoscalar ($g_P$) currents. As the influence of all of the latter is proportional to $q/M \ll 1$, they are often referred to as recoil terms and represent small corrections in the context of allowed $\beta$ decay. Assuming the Conserved Vector Current (CVC) hypothesis \cite{Feynman1958} and the absence of second-class currents \cite{Weinberg1958, Holstein2014a}, of the induced terms all but the weak magnetism current become identically zero up to isospin breaking corrections \cite{Donoghue1990, Paver1991}, and we find additionally
\begin{equation}
    g_M = \kappa_p - \kappa_n = 3.706
\end{equation}
where $\kappa_{p,n}$ is the anomalous magnetic moment of the proton and neutron, respectively. Through CVC and the Ademollo-Gatto theorem $g_V$ remains untouched up to second order in isospin breaking \cite{Ademollo1964}, whereas the axial current is only partially conserved. 

We move towards Eq. (\ref{eq:general_tiple_relation}) by introducing the decay rate for an unpolarized transition
\begin{equation}
    \tau^{-1} = \frac{G_F^2m_e^5}{2\pi^3}|V_{ud}|^2\frac{1}{2J_i+1}\int \mathrm{d}\text{LIPS}\sum_{m_i,m_f}|\mathcal{M}_{fi}|^2,
    \label{eq:general_decay_rate}
\end{equation}
where $\mathcal{M}_{fi}$ is the transition matrix element and $m_{i,f}$ are the projection of initial and final spins along a quantization axis. For a  $J^\pi \to J^\pi$ transition, $\mathcal{M}_{fi}$ receives leading-order contributions from both vector (Fermi, F) and axial-vector (Gamow-Teller, GT) terms when compliant with isospin symmetry.
As a direct consequence one expects $V$-$A$ cross terms in the final transition density. To zeroth order in recoil ($q/M$), however, no such terms appear since the leading vector current term is timelike, while that of the axial current is spacelike. Inspection of Eq. (\ref{eq:nuclear_current}) immediately leads to the usual $\mathcal{M}_F = \langle 1 \rangle$ and $\mathcal{M}_{GT} = \langle \bm{\sigma} \rangle$ reduced matrix elements. For the neutron the situation is particularly simple and one has $\mathcal{M}_F = 1$ and $\mathcal{M}_{GT} = \sqrt{3}$ so that one finds the classical result
\begin{equation}
    \frac{1}{2J_i+1}\sum_{m_i,m_f} |\mathcal{M}_{fi}|^2 = g_V^2(1 + 3 \lambda^2)
\end{equation}
when ignoring higher-order corrections, and where we left out spinor normalizations for clarity. Experimental precision already long ago became sensitive enough to require a theoretical analysis including recoil-order terms, however \cite{Bender1968, Garcia1985, Holstein1974}. Weak magnetism (WM), in particular, shows up at first order in recoil in the spacelike vector current and interferes with the leading Gamow-Teller operator. Collecting terms in the differential spectrum shape one obtains another well-known result
\begin{align}
    \Delta\left(\frac{\mathrm{d}N}{\mathrm{d}W_e}\right)^{\text{WM}} &\propto \frac{4}{3M}\frac{g_M+g_V}{g_A\mathcal{M}_{GT}}p_eW_e(W_0-W_e)^2 \nonumber \\ 
    &\times \left(W_e-\frac{W_0}{2}-\frac{m_e^2}{2W_e}\right)
    \label{eq:weak_magnetism_spectrum_simple}
\end{align}
where $p_e = \sqrt{W_e^2-m_e^2}$ is the electron momentum, $W_e$ its total energy, and $W_0$ the spectrum endpoint. Intriguingly, Weinberg showed \cite{Weinberg1959}, however, that no $V$-$A$ interference terms - such as those in Eq. ($\ref{eq:weak_magnetism_spectrum_simple}$) - can contribute to a scalar quantity such as the decay rate \textit{in the absence of electromagnetic effects}. 
Indeed, performing the integral over Eq. (\ref{eq:weak_magnetism_spectrum_simple}) gives identically zero, leading one to the intuitive conclusion that contributions from Eq. (\ref{eq:weak_magnetism_spectrum_simple}) are suppressed by at least a factor $(\alpha Z)$. In the work by Wilkinson \cite{Wilkinson1982}, the Fermi function is expanded to first order
\begin{equation}
    F(Z, W_e) \approx 1 + \frac{\pi \alpha Z W_e}{p_e},
\end{equation}
and one obtains a non-zero part of the spectrum integral when Eq. (\ref{eq:weak_magnetism_spectrum_simple}) is multiplied by the second term. 
Since $p \approx W$ except for extremely non-relativistic electrons, the total contribution is minimal and one finds a relative contribution on the order of a few $ 10^{-6}$ \cite{Wilkinson1982, Garcia-Luna2006}. Similar results were found for other cross-terms \cite{Wilkinson1982}, resulting in a pure decomposition between vector and axial terms. Taking into account all additional correction factors, the phase space integrals in Eq. (\ref{eq:general_decay_rate}) (denoted by $f_{V,A}$) for both vector and axial terms were found to be equal to a few parts in $10^{6}$. Equation (\ref{eq:general_decay_rate}) then becomes the well-known result
\begin{equation}
    \tau^{-1}_n = \frac{G_F^2g_V^2m_e^5}{2\pi^3}|V_{ud}|^2(1 + 3\lambda^2)f_V(1+RC)
    \label{eq:wilkinson_tau}
\end{equation}
where $f_A/f_V$ is approximated as unity and $f_V = 1.6887(2)$ \cite{Wilkinson1982, Towner2010a}, and incorporating the usual radiative corrections \cite{Sirlin2013}.

We once more take a step back and briefly review the influence of the electromagnetic interaction. Modifications can be summarized into three parts:

(i) static Coulomb interaction between the outgoing $\beta$ particle and the final state and atomic electrons. Part of this is covered by the Fermi function and its higher-order correction terms \cite{Rose1961, Behrens1969, Buhring1984, Hayen2018}.

(ii) additional radiative corrections not contained in (i), e.g. Bremsstrahlung, charge-change and $\gamma W$ boxes \cite{Sirlin2013, Seng2018}, written here as $RC$.

(iii) replacement of $q_0 \to q_0 + e\phi$ where $\phi$ is the electrostatic potential as required by gauge invariance. 

We shall concern ourselves here only with the interplay between (i) and (ii), as (iii) only shows up together with weak magnetism to higher order in recoil.

It was originally shown by Stech and Sch\"ulke that the correct generalization of the transition matrix element in the presence of electromagnetic effects is given by \cite{Stech1964, Holstein1979b}
\begin{align}
    \mathcal{M}_{fi} &= \int \mathrm{d}^3 r\, \bar{\phi}_e(\vec{r}, \vec{p}_e)\gamma^\mu(1+\gamma^5)v(\vec{p}_{\bar{\nu}}) \nonumber \\
    & \times \int \frac{\mathrm{d}^3s}{(2\pi)^3}e^{i\vec{s}\cdot \vec{r}}\frac{1}{2}[\langle f(\vec{p}_f+\vec{p}_e-\vec{s})| V_\mu + A_\mu | i(\vec{p}_i) \rangle \nonumber \\
    & + \langle f(\vec{p}_f) | V_\mu + A_\mu | i(\vec{p}_i-\vec{p}_e+\vec{s}) \rangle ].
    \label{eq:generalized_matrix_element_SS}
\end{align}
where $\bar{\phi}_e$ is the solution to the Dirac equation in the static Coulomb potential of the final state. All form factors are now a function of $q' = (p_f+p_e-s)-p_f$ instead of $q=p_f-p_i$ \cite{Bottino1974}. Inserting Eq. (\ref{eq:nuclear_current}) into this result and taking the weak magnetism term as an example one finds
\begin{align}
    \int &\frac{d^3s}{(2\pi)^3}\left[\int d^3r \, e^{i\vec{r}\cdot \vec{s}} \bar{\phi}_e(\vec{r}, \vec{p}_e)\right]\frac{g_M(q'^2)}{2M} \nonumber \\
    &\times \bar{u}_p\,\sigma_{\mu\nu}[q^\nu + (p_e-s)^\nu]\,u_n\gamma^\mu(1+\gamma^5)v(\vec{p}_{\bar{\nu}}),
    \label{eq:wm_coulomb_general}
\end{align}
where we neglected the nuclear spinor momentum change. It is clear that when $\bar{\phi}_e$ reduces to $\bar{u}(p_e)\,e^{-i\vec{r}\cdot \vec{p}_e}$ when $Z\to 0$, Eq. (\ref{eq:wm_coulomb_general}) evaluates to zero and Eq. (\ref{eq:generalized_matrix_element_SS}) becomes the simple plane wave matrix element. Coulomb corrections with the $q^\nu$ term are $\mathcal{O}(\alpha Z (q/M) qR)$ as anticipated and do not reasonably contribute. The oft-neglected contribution comes from the $(p_e-s)^\nu$ term, which was shown to instead amount to corrections of $\mathcal{O}(\alpha Z / MR)$ \cite{Bottino1974, Holstein1974b}.

The most accurate results are obtained in the Behrens-B\"uhring formalism \cite{Behrens1970, Behrens1982, Stech1964, Schulke1964}, a `calculate first, ask questions later' elementary particle treatment\footnote{The effects of Eq. (\ref{eq:wm_coulomb_general}) were already included by Behrens and B\"uhring before the original publication by Bottino \emph{et al.} \cite{Bottino1974, Holstein1974b} yet appear to have gone unnoticed.}. Collecting terms in the transition matrix element we find the additional correction terms of order $\mathcal{O}(\alpha Z / MR)$ \cite{Hayen2018}
\begin{equation}
    \Delta |\mathcal{M}_{fi}|^2 = \frac{2\sqrt{2}}{3\sqrt{3}}\alpha Z~ {}^VF_{111}^0(1,1,1,1){}^AF_{101}^0
    \label{eq:extra_terms_BB}
\end{equation}
where the ${}^{V/A}F_{KLs}^n(k, m,n,\rho)$ are nuclear form factor coefficients \cite{Behrens1970}. The arguments correspond to the convolution with the nuclear charge distribution, encoded through functions $I(k, m, n, \rho ;r)$ resulting from the expansion of the electron wave function. The shift of Eq. (\ref{eq:extra_terms_BB}) is an energy-independent term in the differential decay cross section proportional to the main Gamow-Teller form factor, denoted by ${}^AF_{101}(q^2)$. Up to higher-order corrections in perturbation theory, the latter can be redefined by incorporating Eq. (\ref{eq:extra_terms_BB}) into an effective GT form factor, thus leaving the usual formulae intact. In the particular case of the neutron this corresponds to a renormalization of $\lambda \to \tilde{\lambda} = \lambda(1+\delta \lambda)$. While in principle similar renormalizations occur for other form factors, these are found to be identically zero due to a cancellation between contributions from the bare charge and the finite size of the charge distribution \cite{Holstein1979}. 
For the neutron, analytic evaluation of Eq. (\ref{eq:extra_terms_BB}) is trivial and we find
\begin{equation}
    \tilde{\lambda} = \lambda \left(1 + \frac{4}{5}\frac{\alpha}{MR}\frac{g_M+g_V}{g_A}\right)^{1/2}
    \label{eq:lambda_renormalization}
\end{equation}
up to $\mathcal{O}((\alpha/MR)^2)$. Using $R = \sqrt{5/3}\langle r^2 \rangle^{1/2}$ for a uniformly charged sphere we find
\begin{equation}
    \delta \lambda = (2.0 \pm 0.1) \cdot 10^{-3}.
    \label{eq:fAfV_neutron}
\end{equation}
where we chose $\langle r^2 \rangle^{1/2} = 0.87(4)$\,fm, taking the uncertainty to be the remaining discrepancy in the proton radius puzzle \cite{Antognini2013, Carlson2015a}.

It is worthwhile to build a qualitative reasoning for this renormalization from a more modern perspective. In writing down Eq. (\ref{eq:generalized_matrix_element_SS}) we have approximately treated electrostatic effects non-perturbatively \cite{Halpern1970, Behrens1982}, roughly corresponding to the exchange of an infinite number of photons between the outgoing $\beta$ particle and initial and final charges. The form of the correction, $\alpha Z / MR$, hints at the underlying structure and points to a single-photon exchange, however. Equation (\ref{eq:extra_terms_BB}) arises due to an interference between vector and axial vector amplitudes with a virtual photon in the final state when squaring the sum of the amplitudes. Naive power counting assumes this interference to be of $\mathcal{O}(\alpha Z (q/M)qR)$, as in the original elementary particle treatment \cite{Bottino1974, Holstein1974b}. One can obtain large remnants in the integration over final state momenta, however. In the triangle diagram with a virtual photon exchanged between the electron and outgoing proton, a term proportional to $\alpha \pi / v_{rel}$ appears, where $v_{rel}$ is the relative velocity between photon and electron. When integrating over the proton energy a non-negligible contribution appears as $v_{rel}$ approaches zero, thereby mitigating the $q/M$ suppression from the original interaction. This will be a topic of further investigation within the context of effective field theories (EFT) \cite{CiriglianoPC2019}.  Additionally, this points to the need for nuance when using power counting arguments in standard EFT approaches, in particular when dealing with virtual photons.

Equation (\ref{eq:lambda_renormalization}) shows that additional radiative corrections to the Gamow-Teller form factor occur on top of the ``inner'' radiative corrections \cite{Marciano2006, Seng2018}. While this is not a problem \textit{an sich}, this reveals two notable issues in the current way of interpreting experimental data for neutron and mirror systems. Both problems stem from the fact that experiment measures the fully renormalized value for $\tilde{\lambda}$ ($\tilde{\rho}$) for the neutron (mirror) system, whatever its value may be. In the experimental analysis, corrections are applied to remove small recoil-order \cite{Holstein1974} and $\mathcal{O}(\alpha)$ radiative corrections \cite{Shann1971, Gudkov2006} specific to the correlation observable. The effect described here is, however, not accounted for in the analysis. Due to differences in the theoretical treatment of neutron and mirrors systems in the literature, one has two distinct consequences.

\begin{figure}[!ht]
    \centering
    \includegraphics[width=0.48\textwidth]{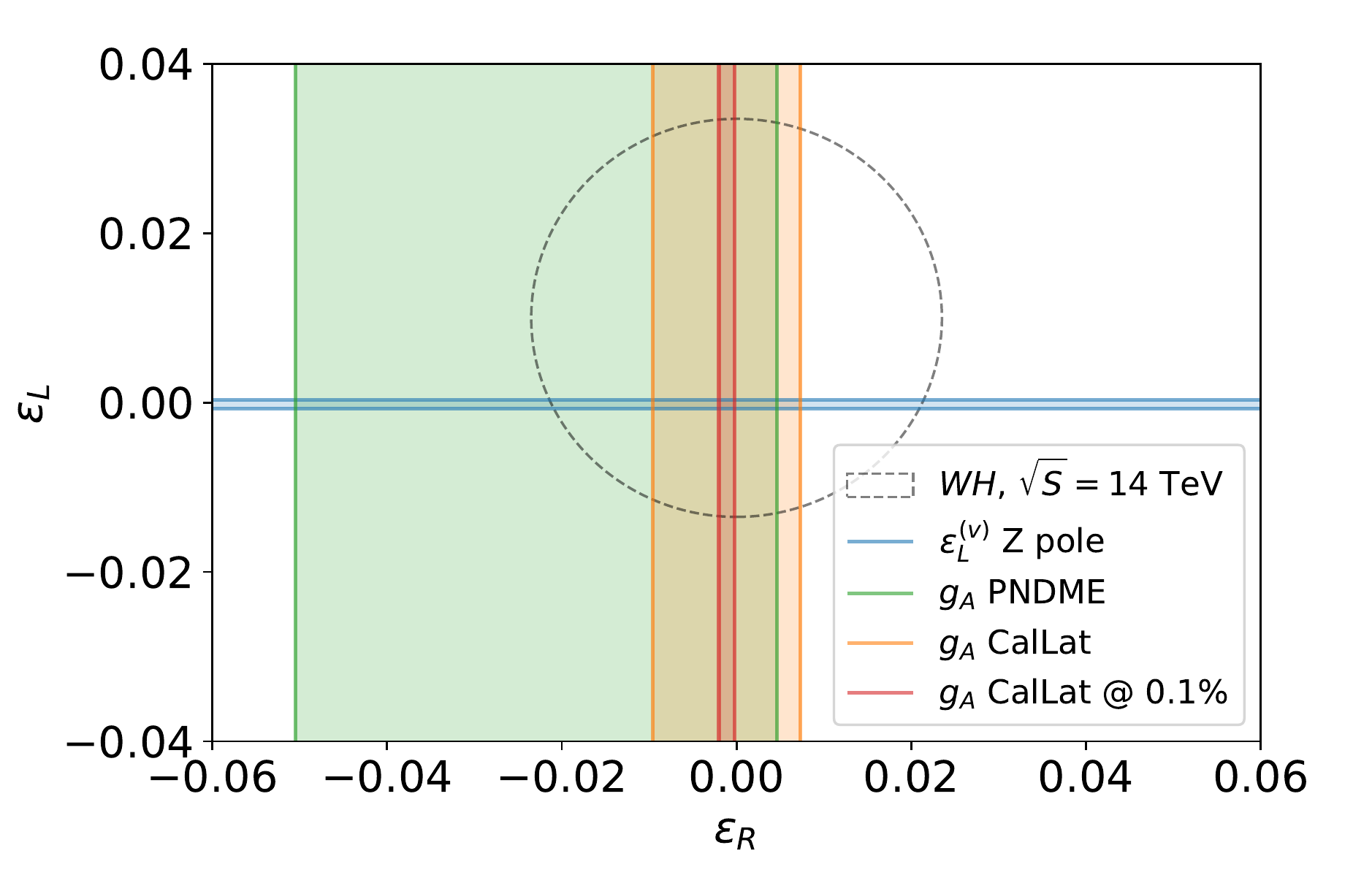}
    \caption{Current limits (90\% C.L.) on left and right-handed couplings interpreted in the SMEFT, showing $Z$-pole (blue) \cite{Falkowski2017, Efrati2015}, LHC (black) \cite{TheATLAScollaboration2016}, and LQCD results from CalLat (orange) \cite{Chang2018} and PNDME (green) \cite{Gupta2018} collaborations. In red we show anticipated limits when $g_A$ reaches 0.1\% on the lattice.}
    \label{fig:epsilonR_limits}
\end{figure}

The first problem presents itself when comparing experimental values for $\tilde{\lambda}$ \cite{Markisch2018, Brown2018, Gonzalez-Alonso2018} to $g_A$ calculated on the lattice \cite{Chang2018, Gupta2018}. Under the usual assumptions, $g_V$ is set equal to unity \cite{Ademollo1964} and all QED effects are contained in $f_V$ and $RC$ of Eq. (\ref{eq:wilkinson_tau}) so that a comparison of experimental and lattice $g_A$ is sensitive to right-handed currents via \cite{Gonzalez-Alonso2018}
\begin{equation}
    \lambda_{EFT} = \lambda_{SM}(1-2~\mathcal{R}e\,[\epsilon_R])
\end{equation}
where $\epsilon_R$ is a BSM right-handed coupling constant assuming new UV physics, interpreted in the Standard Model EFT (SMEFT) \cite{Cirigliano2013a}. The correction derived in Eq. (\ref{eq:lambda_renormalization}) mimics exotic right-handed currents, so that a failure to take it into account would incorrectly lead to a non-zero BSM signal. Figure \ref{fig:epsilonR_limits} shows the corrected current and anticipated limits using $g_A$ from the lattice with the recent experimental PERKEO3 result \cite{Markisch2018}.

The correction corresponds to an 0.1\% shift, putting the current limit at $\epsilon_R = (-1 \pm 5) \cdot 10^{-3}$ when using the CalLat \cite{Chang2018} result. At the current level of precision the effect of the new correction is not significant due to the large uncertainty on $g_A$ from LQCD. Lattice calculations are expected to improve their precision by almost an order of magnitude in the near future, however, meaning Eq. (\ref{eq:lambda_renormalization}) corresponds to a $2\sigma$ shift and becomes of considerable importance. After correcting for Eq. (\ref{eq:lambda_renormalization}), equality between experimental and lattice values for $g_A$ will then put the most stringent direct limits on right-handed currents\footnote{We have omitted here the combination of CKM unitarity ($\Delta_{CKM} \propto \epsilon_L + \epsilon_R$) and the pion decay ($\delta \Gamma_{\pi \to \mu 2} \propto \epsilon_L - \epsilon_R$) due to the degeneracy with pseudoscalar, scalar, and tensor interactions \cite{Alioli2017, Gonzalez-Alonso2018}.}.

The second problem pertains to the evaluation of $V_{ud}$ from mirror decays, i.e. $\beta$ transitions within a $T=1/2$ doublet, analogous to the neutron. The master equation can be obtained by making the substitution $3\lambda^2 \to \rho^2$ in Eq. (\ref{eq:general_tiple_relation}), where $\rho = {}^AF_{101}/{}^VF_{000}$ is the ratio of Gamow-Teller and Fermi form factors. Analogously to the neutron, $\rho$ can be determined experimentally through $\beta(-\nu)$ correlation measurements so that $V_{ud}$ can be extracted given theoretical calculations for $f_A/f_V$. Due to an apparent divergence in the literature, the problem is opposite to that of the neutron as the effect of Eq. (\ref{eq:extra_terms_BB}) was explicitly included in the $f_A/f_V$ ratio \cite{Severijns2008, Towner2015}. Since the analysis of experimental data returns $\tilde{\rho}$ - which includes the renormalization described here - its incorporation in $f_A/f_V$ then results in a double-counting. We recalculate the reported $f_A/f_V$ values \cite{Severijns2008, Naviliat-Cuncic2009} for the isotopes for which all experimental information is available to allow extraction of $V_{ud}$: $^{19}$Ne, $^{21}$Na, $^{29}$P, $^{35}$Ar and $^{37}$K. Results are listed in Table \ref{tab:new_Vud_mirrors}.

\begin{table}[!ht]
    \centering
    \begin{ruledtabular}
    \begin{tabular}{r|llll}
    & $(f_A/f_V)^\text{old}$ & $(f_A/f_V)^\text{new}$ & $\mathcal{F}t_0^\text{old}$ & $\mathcal{F}t_0^\text{new}$ \\
    \hline 
    $^{19}$Ne \cite{Calaprice1975a} & 1.0143(29) & 1.0012(2) & 6189(28) & 6131(25)\\
    $^{21}$Na \cite{Vetter2008} & 1.0180(36) & 1.0019(4) & 6185(44) & 6152(42)\\
    $^{29}$P \cite{Masson1990} & 1.0223(45) & 0.9992(1) & 6535(606) & 6496(593)\\
    $^{35}$Ar \cite{Naviliat-Cuncic2009} & 0.9894(21) & 0.9930(14) & 6133(51) & 6135(51)\\
    $^{37}$K \cite{Shidling2014, Fenker2018} & 1.0046(9) & 0.9957(9) & 6148(33) & 6135(33)
    \end{tabular}
    \end{ruledtabular}
    \caption{Difference in calculated $f_A/f_V$ values and its effect on $\mathcal{F}t_0$ for the mirror $T=1/2$ transitions for which all experimental information is available to allow extraction of $|V_{ud}|$. $\mathcal{F}t$ value are taken from \cite{SeverijnsTBP} for all isotopes.}
    \label{tab:new_Vud_mirrors}
\end{table}

\begin{figure}[!ht]
    \centering
    \includegraphics[width=0.48\textwidth]{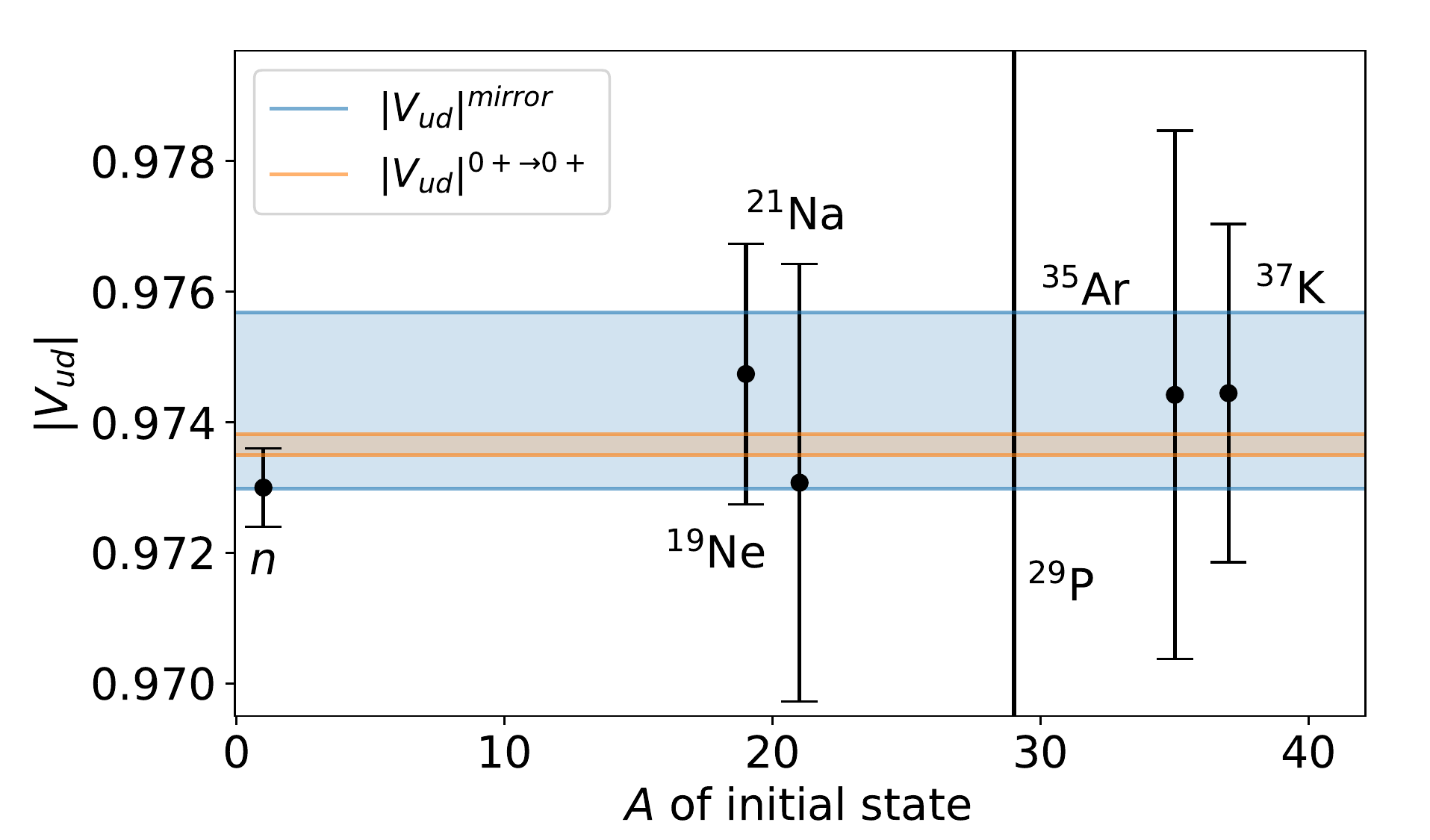}
    \caption{Results with $1\sigma$ uncertainty of $|V_{ud}|$ from nuclear decays. The new correction greatly improves internal consistency among mirror decays and pushes $|V_{ud}|^\text{mirror}$ upwards to within $1\sigma$ of superallowed decays.}
    \label{fig:overview_Vud_nuclear}
\end{figure}

As discussed above, the largest difference in $f_A$ and $f_V$ comes from the constant term of Eq. (\ref{eq:extra_terms_BB}) due to the factor $\alpha Z$ suppression of Eq. (\ref{eq:weak_magnetism_spectrum_simple}). As a consequence, differences are now much smaller as finite size corrections \cite{Hayen2018} are very similar for axial and vector transitions. Like in Ref. \cite{Naviliat-Cuncic2009}, we have assumed a $20\%$ uncertainty on the deviation from unity for $f_A/f_V$. It serves as an input to the corrected $ft$ value common to all mirror decays, $\mathcal{F}t_0$, which is defined as \cite{Naviliat-Cuncic2009}
\begin{align}
    \mathcal{F}t_0 &= g_V^2f_Vt(1+\delta_R^\prime)(1+\delta^V_{NS}-\delta_C^V) [1+(f_A/f_V)\rho^2] \nonumber \\
    &\equiv \mathcal{F}t[1+(f_A/f_V)\rho^2],
\end{align}
where $\delta_i$ are additional radiative ($R$), nuclear structure ($NS$) and isospin-breaking ($C$) corrections \cite{Severijns2008}. The change in $\mathcal{F}t_0$ is strongest for $^{19}$Ne due to the large value for $\rho$. Combining all newly calculated results, one obtains an average $\overline{\mathcal{F}t}_0 = 6136(17)$ with $\chi^2/\nu = 0.14$, resulting from the significantly enhanced internal consistency. One then relates this to $|V_{ud}|$ via \cite{Naviliat-Cuncic2009, Naviliat-Cuncic2013}
\begin{equation}
    V_{ud}^2 = \frac{K}{\overline{\mathcal{F}t}_0G_F^2(1+\Delta_R^V)}
    \label{eq:Vud_mirrors_general}
\end{equation}
where $K/(\hbar c)^6 = 2\pi^3\, \ln{2\hbar}/(m_ec)^5 = 8120.278(4) \times 10^{-10}\,$GeV$^{-4}\,$s, $G_F/(\hbar c)^3 = 1.1663787(6) \times 10^{-5}\,$GeV$^{-2}$ is the Fermi coupling constant \cite{Tishchenko2013} and $\Delta_R^V = 2.467(22)\%$ is the so-called inner radiative correction \cite{Marciano2006, Seng2018}. Application of Eq. (\ref{eq:Vud_mirrors_general}) then leads to a new value for $|V_{ud}|$ extracted from mirror decays
\begin{equation}
    |V_{ud}|^\text{mirror} = 0.9743(14)
\end{equation}
which lies $0.3\%$ ($2.2\sigma$) higher than the results previously reported \cite{Naviliat-Cuncic2009, Naviliat-Cuncic2013} when using the new radiative corrections as above \cite{Seng2018}, $|V_{ud}|^\text{mirror}_\text{old} = 0.9712(14)$. This new result agrees well with that of superallowed Fermi decays, $|V_{ud}|^{0+\to 0+} = 0.97366(16)$ \cite{Hardy2015, Seng2018}. Figure \ref{fig:overview_Vud_nuclear} shows an overview of the current status.

In conclusion, we have discussed an additional radiative correction which arises only for Gamow-Teller allowed decays and results from an interplay between QED and QCD-induced currents. In the case of the neutron this emerges as an additional renormalization of $g_A$ not contained in the usual radiative corrections. We have shown this effect to be essential in the search for Beyond Standard Model physics when comparing to results of LQCD when the precision of the latter significantly improves in the near future. In the mirror systems, where a precise knowledge of theoretical corrections is analogously used for the extraction of $V_{ud}$, we have shown that this additional correction is double-counted when combined with experimental data. We have corrected for this and demonstrated the resulting increased consistency within the mirror data set, and derived a new value for $V_{ud}$. The resultant $|V_{ud}|$ value is 0.3\% higher than the previous result and is consistent with that of superallowed Fermi decays.

\begin{acknowledgements}
One of the authors (L.H.) would like to thank Albert Young for the inspiration to look into this material. Further, the authors would like to thank Vincenzo Cirigliano, Mikhail Gorchtein, Barry R. Holstein and the organizers of ECT*: \textit{Precise beta decay calculations for searches for new physics} and ACFI Amherst: \textit{Current and Future Status of the First-Row CKM Unitarity} workshops for productive discussions related to this manuscript. This work has been partly funded by the Belgian Federal Science Policy Office, under Contract No. IUAP EP/12-c and the Fund for Scientific Research Flanders (FWO).
\end{acknowledgements}

\bibliography{library}
\end{document}